\documentclass{appolb}
\usepackage{graphicx}
\usepackage{hyperref} 

\begin{document}
\title{Mixture of quark and gluon fluids described in terms of anisotropic hydrodynamics%
\thanks{Presented at the 9th workshop in a serie ,,Excited QCD 2017'', Sintra, Lisbon, Portugal, May 7-13, 2017.}%
}
\author{E. Maksymiuk
\address{Institute of Physics, Jan Kochanowski University, PL-25-406 Kielce, Poland}
}
\maketitle
\begin{abstract}
A system of equations of anisotropic hydrodynamics that describes mixture of quark and gluon fluids is studied. The equations are based on the zeroth, first, and second moments of the RTA kinetic equations. Tests of this formulation are performed by comparing the results of anisotropic hydorodynamics with the exact solutions of the Boltzmann equations for a mixture of fluids in the Bjorken flow limit. One finds a very good agreement between the hydrodynamic and kinetic-theory results \cite{Florkowski:2015cba}. 
\end{abstract}
\PACS{PACS numbers come here}
 
\section{Introduction}
%
Relativistic viscous hydrodynamics has been used as a fundamental tool to understand the evolution of matter produced in heavy-ion experiments at RHIC and the LHC~\cite{Florkowski:2010zz,Heinz:2013th, Betz:2008me,  Denicol:2010xn, Martinez:2010sc, Florkowski:2010cf,Strickland:2014pga}.  Despite the success of traditional viscous hydrodynamics in reproducing collective behavior 
of matter, there are still theoretical shortcomings that may question the validity of such an approach in heavy-ion experiments conditions. Large flow gradients and fast longitudinal expansion produce very large pressure corrections, in contrast to the founding hydrodynamic hypotheses of small deviations from local equilibrium and perturbative treatment of viscous corrections. One way to address this problem is anisotropic hydrodynamics~\cite{Tinti:2013vba,Florkowski:2014bba,Nopoush:2015yga,Ryblewski:2010tn}.

Most of theoretical investigations on relativistic hydrodynamics start with a kinetic theory and this is also the case for studies of mixtures~\cite{Florkowski:2013lya,Florkowski:2014sfa,Florkowski:2015lra,Denicol:2014xca}. However, a very good agreement of anisotropic hydrodynamics with the exact solutions of the Boltzmann equations, found for simple fluids, has not been confirmed in early works on mixtures~\cite{Florkowski:2012as,Florkowski:2014txa}. This suggests using a more general approach than that
presented in~\cite{Florkowski:2012as,Florkowski:2014txa}, which is reported in this paper.

\section{Kinetic equations}
%
We start our analysis with the kinetic equations for quarks, antiquarks and gluons written in the relaxation time approximation (RTA)~\cite{Bhatnagar:1954zz,aw,book}
\begin{equation}
 p^{\mu }\partial_{\mu } f_i (x,p)= 
- p^\mu U_\mu \frac{f_i(x,p) - f_{i, {\rm eq}}(x,p)}{\tau_{\rm eq}},  
\label{kine0}
\end{equation}
where $i$ corresponds to $Q^+$, $Q^-$ or $G$,  $f_i$ is phase-space distribution function, and $\tau_{\rm eq}$ is the relaxation time.

The quark and gluon distribution functions are assumed to have a generic structure~\cite{Romatschke:2003ms}
\begin{eqnarray}
f_{Q^\pm}(x,p) &=& \exp\left(\frac{\pm \lambda - \sqrt{(p\cdot U)^2 + \xi_q (p\cdot Z)^2}}{\Lambda_q} \right),
\\
f_G(x,p) &=& \exp\left(-\frac{\sqrt{(p\cdot U)^2 + \xi_g (p\cdot Z)^2}}{\Lambda_g} \right),
\label{RSform}
\end{eqnarray}
where $\Lambda_q$ and $\Lambda_g$ define the transverse momentum scale, $\lambda$ is the non-equilibrium baryon chemical potential of quarks, while $\xi_q$ and $\xi_g$ are the anisotropy parameters. Moreover $U^\mu = (t,0,0,z)/\tau$ and  $Z^\mu = (z,0,0,t)/\tau$,  where $\tau =\sqrt{t^2-z^2}$ is the longitudinal proper time.

In the local equilibrium, the two anisotropy parameters vanish, $\Lambda_q$ and $\Lambda_g$ become equal to $T$,  and $\lambda$ becomes $\mu$, namely
\begin{eqnarray}
f_{Q^\pm,{\rm eq}}(x,p) = \exp\left(\frac{\pm \mu - p\cdot U}{T}  \right),
\quad
f_{G,{\rm eq}}(x,p) = \exp\left(-\frac{p\cdot U}{T}  \right).
\label{eqforms}
\end{eqnarray}
The equilibrium distribution functions are used to define the RTA collision terms in~(\ref{kine0}). In this case $\mu$ and $T$ should be treated as the effective baryon chemical potential and effective temperature that are determined by the appropriate Landau matching conditions. For simplicity, we assume here the classical Boltzmann statistics. 

\section{Moments of the kinetic equations}
%
In this section we introduce equations of anisotropic hydrodynamics. This is done by using moments of the kinetic equations. 
%
\subsection{Zeroth moments of the kinetic equations}
Integrating Eq.~(\ref{kine0}) over three-momentum and including the internal degrees of freedom we obtain the three scalar equations
\begin{eqnarray}
\partial_{\mu } (n_{i} U^{\mu}) &=& 
 \frac{n_{i, {\rm eq}} - n_{i}}{\tau_{\rm eq}}, 
\label{den}
\end{eqnarray}
where we have introduced the non-equilibrium and equilibrium particle densities, see Ref.~\cite{Florkowski:2015cba}.
Instead of using Eq.~(\ref{den}) we use the difference of the equations for quarks and antiquarks appearing in (\ref{den}). Dividing it by a factor 3 gives the constraint on the baryon number density.
Using baryon number conservation we have found $\lambda_q/\Lambda_q$ and $\mu/T$ defined by the functions
\begin{eqnarray}
D(\tau,\Lambda_q,\xi_q) =  \left(\frac{3 \pi^2 b_0 \tau_0 \sqrt{1+\xi_q}}{2 g_q \tau \Lambda_q^3} \right), 
\quad
\kappa_q(T,\Lambda_q,\xi_q) = \frac{T^3  \sqrt{1+\xi_q}}{\Lambda_q^3}.
\label{D}
\end{eqnarray}
Using this notation we write first equation of anisotropic hydrodynamics
\begin{eqnarray}
&&\frac{d}{d\tau}  \left(\alpha \frac{ \sqrt{1+D^2}  \Lambda_q^3}{\sqrt{1+\xi_q}} 
+ (1-\alpha)   \frac{{\tilde r} \Lambda_g^3}{\sqrt{1+\xi_g}} \right)
\nonumber \\
&&+  \left( \frac{1}{\tau} + \frac{1}{\tau_{\rm eq}} \right) 
\left( \alpha \frac{\sqrt{1+D^2}  \Lambda_q^3}{\sqrt{1+\xi_q}} + (1-\alpha)  \frac{ {\tilde r}\Lambda_g^3}{\sqrt{1+\xi_g}}  \right)
\nonumber \\
&&  = \frac{T^3}{\tau_{\rm eq}}  \left( \alpha \sqrt{1+ D^2/\kappa_q^2}+ (1-\alpha) {\tilde r} \right),
\label{3zero}
\end{eqnarray}
where ${\tilde r} = 2/3 $ is the ratio of the quark and gluon internal degrees of freedom. The parameter $\alpha$ should be taken from the range $0 \leq \alpha \leq 1$, see Ref.~\cite{Florkowski:2015cba}.

\subsection{First moments of the kinetic equations}
%
The energy-momentum conservation law for the system of partons has the form $\partial_\mu T^{\mu\nu} = 0$.  
Landau matching condition for the energy-momentum conservation requires that the energy determined from  the non-equilibrium distribution functions is the same as the energy obtained with the equilibrium distribution functions
$
\varepsilon = \varepsilon_q + \varepsilon_g = \varepsilon_{\rm eq} = \varepsilon_{q, \rm eq} + \varepsilon_{g, \rm eq}.
$
This leads directly to the constraint on the effective temperature $T$,
\begin{eqnarray}
T^4=  \, \frac{ \Lambda^4_q \sqrt{1+D^2} \, {\cal R}(\xi_q)  + \Lambda^4_g{\tilde r} 
\, {\cal R}(\xi_g) }{ \sqrt{1+D^2/\kappa_q^2} + {\tilde r}},
\label{TL}
\end{eqnarray}
with function ${\cal R}(\xi)$ defined in~\cite{Martinez:2010sc}. In the (0+1)D case considered here the energy and momentum conservation takes the form 
\begin{equation}
\frac{d\varepsilon}{d\tau} = -\frac{\varepsilon+P_L}{\tau},
\label{enmomcon01}
\end{equation}
where $P_L$ is the sum of the longitudinal pressures for quarks and gluons.
This leads directly to the formula
\begin{eqnarray}
&&\frac{d}{d\tau} \left[    \Lambda_q^4 \sqrt{1+D^2} \,
{\cal R}(\xi_q)  + {\tilde r} \Lambda_g^4 {\cal R}(\xi_g) \right] \label{eneq} \\
&& = \frac{2}{\tau} \left[ \Lambda^4_q \sqrt{1+D^2} \, (1+\xi_q) 
{\cal R}^\prime (\xi_q)  + {\tilde r}  \Lambda^4_g (1+\xi_g)  {\cal R}^\prime (\xi_g) \vphantom{e^{\lambda/\Lambda}} \right],
 \nonumber
\end{eqnarray}
with ${\cal R}_L$ defined in~\cite{Martinez:2010sc}.

\subsection{Second moments of the kinetic equations}
%

Second moment of the Boltzmann equation was studied in detail in Ref.~\cite{Tinti:2013vba}. In our one-dimensional case only one of three equations selected as the basis for the momentum anisotropy is independent. It may be taken as
\begin{eqnarray}
\frac{d}{d\tau}\ln\Theta_X-\frac{d}{d\tau}\ln\Theta_Z-\frac{2}{\tau}=\frac{\Theta_{\rm eq}}{\tau_{\rm eq}}\left[\frac{1}{\Theta_X}-\frac{1}{\Theta_Z}  \right].
\label{sumX}
\end{eqnarray}
Following the method of Ref.~\cite{Florkowski:2014bba} one can derive the formulas for $\Theta$ functions for quarks and gluons, and close the system of anisotropic hydrodynamics equations with the following equations
\begin{eqnarray}
&&\frac{d}{d\tau}\ln\left(\frac{\Lambda_q^5}{(1+\xi_q)^{1/2}}\sqrt{1+D^2}\right) 
\label{Tintiq} 
-\frac{d}{d\tau}\ln\left(\frac{\Lambda_q^5}{(1+\xi_q)^{3/2}}\sqrt{1+D^2}\right)-\frac{2}{\tau} 
\nonumber \\
&&= \frac{T^5}{\tau_{\rm eq}\Lambda^5_q}\xi_q(1+\xi_q)^{1/2}\frac{\sqrt{1+D^2/\kappa_q^2}}{\sqrt{1+D^2}},
\nonumber 
\end{eqnarray}
and 
\begin{eqnarray}
&&\frac{d}{d\tau}\ln\left(\frac{\Lambda_g^5}{(1+\xi_g)^{1/2}}\right) 
- \frac{d}{d\tau}\ln\left(\frac{\Lambda_g^5}{(1+\xi_g)^{3/2}}\right)-\frac{2}{\tau} \
 = \frac{T^5}{\tau_{\rm eq}\Lambda^5_g}\xi_g(1+\xi_g)^{1/2}. \nonumber\\
\label{Tintig}
\end{eqnarray}

\section{Results}
%
\begin{figure}[t!]
\includegraphics[angle=0,width=0.5\textwidth]{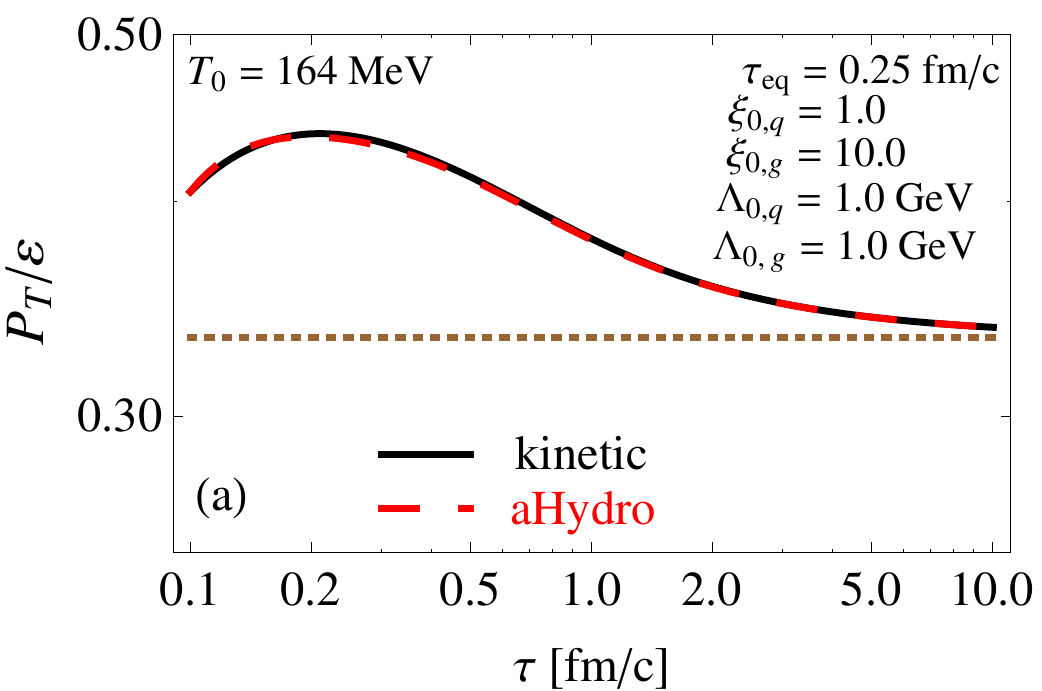} 
\includegraphics[angle=0,width=0.5\textwidth]{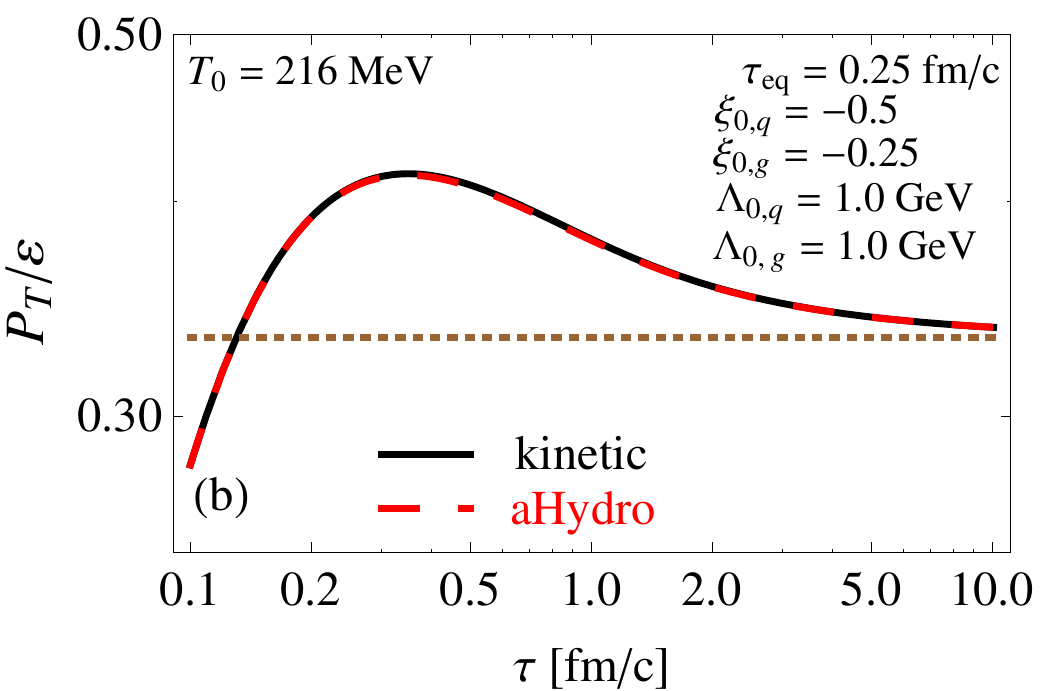}  \\
\includegraphics[angle=0,width=0.5\textwidth]{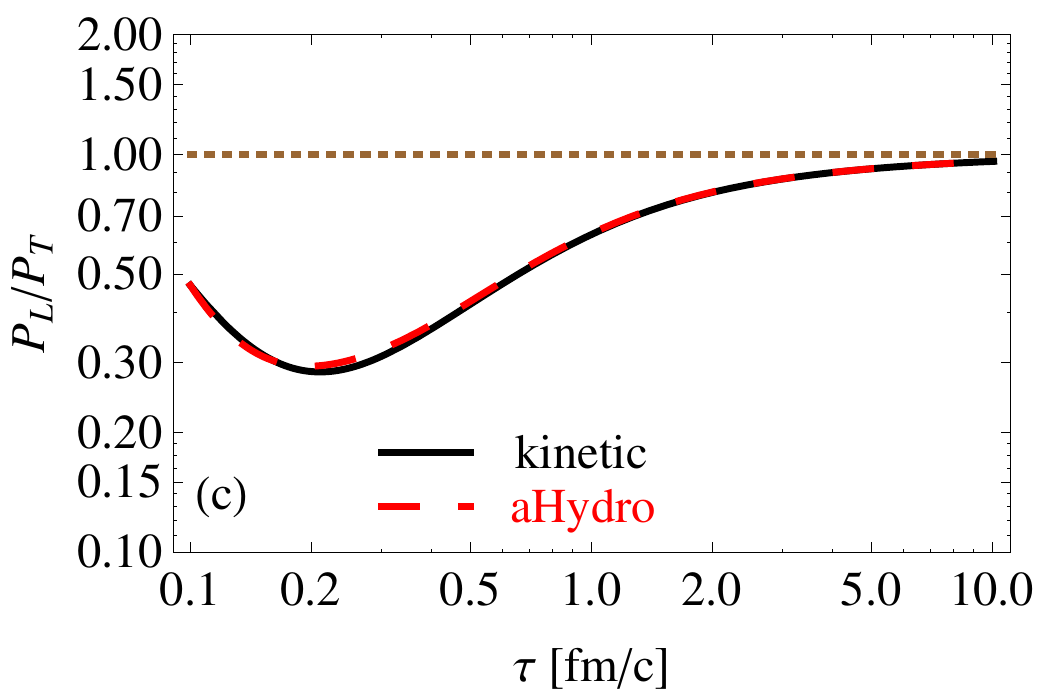} 
\includegraphics[angle=0,width=0.5\textwidth]{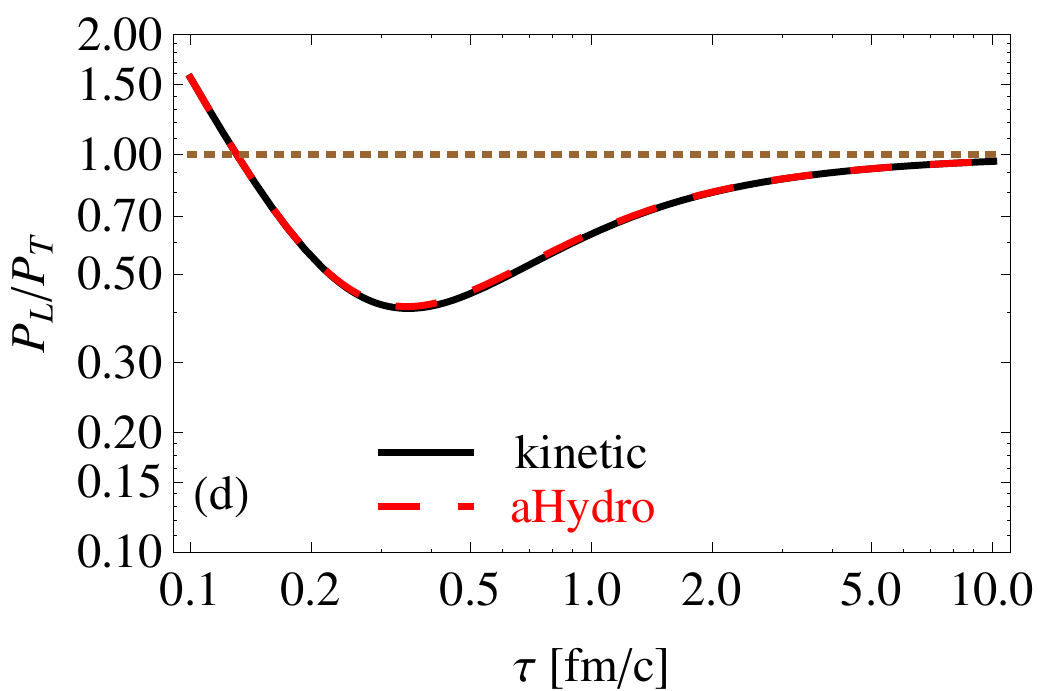}  \\
\caption{Comparison of the anisotropic hydrodynamics  and  kinetic-theory results for the initial oblate-oblate (a)--(c) and prolate-prolate (b)--(d) configurations. In~these calculations parameter $\alpha=1$.}
\label{fig:results}
\end{figure}
Our numerical results presented in this section include two types of initial conditions. Figs.~\ref{fig:results}a and~\ref{fig:results}c correspond to the oblate quark and gluon distribution functions (where the two anisotropy parameters $\xi_q$ and $\xi_g$ are positive and the transverse pressure is larger then the longitudinal one), while Figs.~\ref{fig:results}b and~\ref{fig:results}d present  two initially prolate distribution functions ($\xi_q$ and $\xi_g$ parameters are negative and transverse pressure is smaller then longitudinal one).

Expansion considered in this paper starts at the proper time $\tau_0=0.1$~fm/c and is continued till $\tau=$~10~fm/c. The relaxation time is constant, $\tau_{\rm eq} = 0.25$~fm/c. The initial transverse-momentum parameters $\Lambda_i(\tau_0)$ for quarks and gluons have been set equal to 1~GeV.

Figure~\ref{fig:results} presents a comparison between numerical results obtained from the kinetic theory (black lines) and anisotropic hydrodynamics (red lines). Exact solutions of the Boltzmann equation for (0+1)D systems were constructed earlier in Ref.~\cite{Florkowski:2014txa}. The results presented here include the ratios of the total transverse pressure to the total energy density, $P_T/\varepsilon$, and of the total longitudinal pressure to transverse pressure, $P_L/P_T$.

We have found a good agreement between kinetic theory and anisotropic hydrodynamics. Our results agree with the expectation that the ratio $P_T/\varepsilon$ should be equal to $1/3$ in thermodynamic equilibrium. Similarly, the transverse and longitudinal pressures are almost equal for the late time of the evolution. 

\section{Summary}
%
Using the zeroth, first, and the second moments of the RTA kinetic equations we closed the set of equations for anisotropic hydrodynamics for a mixture of quark and gluon fluids. In a contrast to previous studies, based only on the zeroth and first moments, a very good agreement between kinetic theory and anisotropic hydrodynamics for initially oblate-oblate and prolate-prolate systems has been found. 

%
\section*{Acknowledgments}
%
I would like to thank Wojciech Florkowski and Radoslaw Ryblewski for clarifying and useful discussions.
%

\end{document}